\documentclass[article]{JHEP3} 

\usepackage{epsfig,multicol,amsmath}

\newcommand\fverb{\setbox\pippobox=\hbox\bgroup\verb}
\newcommand\fverbdo{\egroup\medskip\noindent%
			\fbox{\unhbox\pippobox}\ }
\newcommand\fverbit{\egroup\item[\fbox{\unhbox\pippobox}]}
\newbox\pippobox

\title{Rotating Supertubes}
\author{Jin-Ho Cho and Phillial Oh\\
	BK21 Physics Research Division and Institute of Basic Science\\
 Sungkyunkwan University, Suwon 440-746, Korea\\
	E-mail: \email{jhcho@taegeug.skku.ac.kr}, \email{ploh@newton.skku.ac.kr}}

\preprint{\hepth{0302172}}	

\abstract{We study the rotating tubular D2-brane as a time dependent supersymmetric solution of type-IIA string theory. We show that the Poynting angular momentum of the supertube can be replaced by the mechanical angular momentum without disturbing the 8 supersymmetries. Unlike the non-rotating supertube, whose cross section can take an arbitrary shape, the rotating supertube admits only the circular cross section. When there is no electric field on the world-volume, the supersymmetry dictates the angular velocity of the tubular D2-brane to be inversely proportional to the magnetic field. This rotating supertube can be considered as the `blown-up' configuration of an array of spinning D0-particles and is T-dual to the spiraling D-helix whose pitch moves at the speed of light.}


\begin{document} 

\section{Introduction}


The experimental evidence for the positive cosmological constant \cite{riess} calls for our attention to the time dependent solutions in superstring theory \cite{strominger,sen,costa,wang}. These solutions involve the brane configurations which are unstable by construction, therefore not supersymmetric and usually have tachyonic modes on the world-volumes of the branes. (See Ref. \cite{sen2} for example.) 

On the other hand recently, a {\it time dependent but supersymmetric} solution was found in Ref. \cite{townsend}. The solution describes a tubular D2-brane named as `supertube' carrying some angular momentum. Although the supersymmetry algebra $\{Q,\,Q\}\sim {\cal H}$ seems to imply the inconsistency between the time dependency and the supersymmetry in general, the supertube is a completely consistent solution. This is because the angular momentum is the Poynting vector provided by the Born-Infeld (BI) gauge fields on its world-volume and only the gauge connection component $A_x$ is time dependent \cite{townsend2}. By taking T-duality on this supertube, one can obtain another interesting time dependent supersymmetric solution, D-helix \cite{choh}, that is a coiled D-string in a real motion along the axial direction at the speed of light. (See Ref. \cite{lunin} for another derivation of the solution.) This solution is also consistent with the supersymmetry because its (time dependent) position in the embedding flat spacetime geometry does not specify the state of D-helix. 

Since we are lacking of an effective tool for the theory involving tachyons, the time dependent supersymmetric solutions could provide good toy models to build up our intuition for the time dependent systems, and could possibly be staring points in the arena of the aforementioned stringy cosmology. In fact, for the problems of the spacelike (or null) singularity, the time dependent supersymmetric solutions have been explored to see whether they provide the stable backgrounds amenable for exact perturbative string analysis \cite{seiberg,cornalba,seiberg2,horowitz,figueroa,bachas,fabinger}. Along this line, several time dependent systems of supersymmetric intersecting D-strings were analyzed in type-IIB theory \cite{bachas,myers,cho,bachas2}. 

On the IIA side, an interesting time dependent solution of a rotating ellipsoidal D2-brane was constructed in Ref. \cite{savvidy1}. Although the solution is classically stable \cite{savvidy2}, it is not supersymmetric. In this paper, we consider a rotating tubular D2-brane as a possible time dependent supersymmetric configuration that is in a real motion. We recall that BPS Q-balls or some topological solitons are stabilized due to the angular momentum \cite{leese,ward}. In the non-rotating supertube case, the angular momentum is given by the BI gauge fields.

We show that the Poynting angular momentum of the supertube can be replaced by the mechanical angular momentum without disturbing the 8 supersymmetries. Unlike the non-rotating supertube, whose cross section can take an arbitrary shape \cite{bak,mateos}, the rotating supertube admits only the circular cross section. When there is no electric field on the world-volume, the supersymmetry dictates the angular velocity of the tubular D2-brane to be inversely proportional to the magnetic field. This rotating supertube can be considered as the `blown-up' configuration of an array of spinning D0-particles and is T-dual to the spiraling D-helix whose pitch moves at the speed of light. 

The paper is organized as follows. In Sec. \ref{ii}, we construct BI action for a rotating tubular D2-brane and obtain its equations of motion. In Sec. \ref{iii}, we analyze the supersymmetry for the rotating tube and show that 8 supersymmetries are preserved regardless of the rotation. Especially we use a specific representation for the spinors and find out the supersymmetric conditions explicitly. In the subsequent subsections, we apply these supersymmetric conditions to the equations of motion and find the solutions. For the non-rotating case, we recover the results discovered in Ref. \cite{townsend,mateos,cho1} and obtain the equation governing the supersymmetric profile of the radius varying configuration (Sec. \ref{iiia}). For the rotating case, the supersymmetric condition allows only circular cross section of the configuration (Sec. \ref{iiib}). Without any electric field assumed over the world-volume of the tubular D2-brane, the rotation of the magnetic flux induces the electric field over the configurations and stabilizes them. As for the BIon spike solution, the mechanical rotation of the planar D2-brane carrying magnetic flux triggers the electric charge. In Sec. \ref{iv}, we show that the tensionless conditions agree with the supersymmetric conditions. In the rotating case, the tension around the angular direction vanishes only when the cross section takes the circular shape. In Sec. \ref{v}, we discuss about the BPS bound of the Hamiltonian and argue that the charge concerned with the rotation is due to the fundamental strings induced under the moving magnetic flux. In Sec. \ref{vi}, we discuss about the angular momentum bound which forbids the supertube to rotate with the superluminal linear speed. The bound tells us that the charge density of the induced fundamental strings cannot exceed that of the D0-branes dissolved on the tube. Sec. \ref{vii} is devoted to a T-dual configuration of the rotating supertube. Taking T-duality along the axial direction of the rotating supertube, we obtain the spiraling D-helix, whose pitch moves at the speed of light. Sec. \ref{viii} concludes the paper with some discussions on the M-brane configuration of the rotating supertube. We also confirm the circular shape of the rotating supertube by showing (in the Appendix) that two tilted D-strings passing by each other cannot preserve any supersymmetry even when the intersection point moves at the speed of light. Lastly we give some remarks on the zero radius of the rotating supertube.   

\section{A Tubular D2-brane in the Rotational Motion}\label{ii}

Let us consider a tubular D2-brane embedded in a trivial type-IIA background, that is, with no dilaton field, NS-NS $B$ field, and with flat geometry;
\begin{eqnarray}\label{metric}
ds^2=-dT^2+dX^2+R^2d\Phi^2+dR^2+ds^2(E\!\!\!\!E^{(6)}).
\end{eqnarray} 
In order to allow the rotational motion of the tube, we embed the world-volume angular coordinate $\varphi$ into the above flat spacetime background in a time dependent manner; $\Phi=\Phi(t,\,\varphi)$. The BI action has the world-volume reparametrization symmetry. For our purpose, it is convenient to choose the following `comoving gauge' for the world-volume coordinates $(t,\,x,\,\varphi)$;
\begin{eqnarray}\label{gauge}
dT=dt,\quad dX=dx,\quad d\Phi=d\varphi+\Phi_t dt.
\end{eqnarray}
The world-volume coordinate frame is therefore rotating with respect to the spacetime coordinate frame with the angular velocity, $\Phi_t\equiv \partial\Phi/\partial t$ and the coordinate field $\Phi(t)$ will be considered as one of the dynamical variables for the system. {\it In the above specific gauge}, we let the radial profile $R$ of the tube vary with the coordinates $(x,\, \varphi)$ but not with time $t$. In other words, the tubular D2-brane has a solid radial profile (in the world-volume coordinates) but is rotating with respect to the spacetime coordinate frame. The geometry induced on the tube becomes
\begin{eqnarray}\label{metric2}
ds^2&=&g_{\alpha\beta}d\sigma^\alpha d\sigma^\beta\\
&=&-\left(1-R^2\Phi_t^2\right)dt^2+\left(1+R_x^2\right)dx^2+\left(R^2+R_\varphi^2\right)d\varphi^2+2R_xR_\varphi dx\,d\varphi+2R^2\Phi_t\,dt\,d\varphi,\nonumber
\end{eqnarray}
where $R_x\equiv\partial R/\partial x$ and $R_\varphi\equiv \partial R/\partial\varphi$.
We consider the following BI 2-form field strengths {\it in the above specific gauge} (\ref{gauge}):
\begin{eqnarray}\label{BIfield} 
F=E\,dt\wedge dx+B\,dx\wedge d\varphi.
\end{eqnarray} 
We absorbed the scale $2\pi l_s^2$ into the the above definition so that the field $E$ is dimensionless while the component $B$ has the dimension of length.
Note that the electric field $E$ and the magnetic field $B$ are defined in the comoving frame. Actually this setup of the fields gives the true meaning of the `rotation' of the world-volume. The magnetic flux, written in the rotating world-volume coordinates as 
\begin{eqnarray}\label{flux}
\int\limits dx \,d\varphi\,\, B,
\end{eqnarray}
is due to the D0-branes dissolved over the world-volume of the tubular D2-brane. The electric field $E$ is produced by the end points of the macroscopic strings laid along the axial direction of the tube. Therefore we arranged D0-branes and macroscopic strings so that they be stalled with respect to the world-volume frame of Eq. (\ref{gauge}) but rotate with respect to the spacetime coordinate frame. The whole situation is depicted in Fig. 1.

\epsfbox{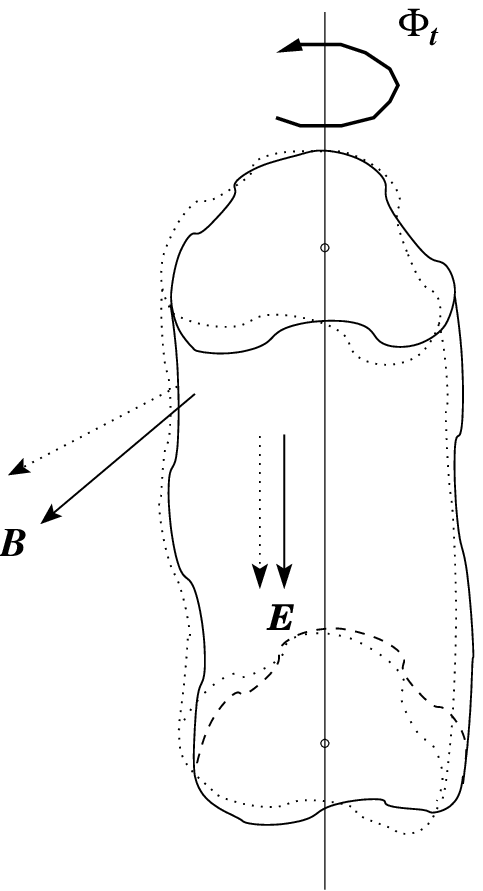}
\begin{flushleft}
{\footnotesize Fig. 1: The tubular D2-brane is in the rotational motion. The magnetic flux and the electric field are stalled on the tube and are rotating with respect to the spacetime coordinates. 
}
\end{flushleft}

The BI Lagrangian for the tube will be
\begin{eqnarray}\label{lagrangian}
{\cal L}
&=&-\frac{1}{g_sl_s(2\pi l_s)^2}\sqrt{-\det\left(g+F\right)}\nonumber\\
&=&-\frac{1}{g_sl_s(2\pi l_s)^2}\sqrt{R^2\left[1-\left(E+\Phi_t B\right)^2\right]+B^2+\left(1-E^2-R^2\Phi_t^2\right)R_\varphi^2+R^2R_x^2}\nonumber\\
&\equiv&-\frac{1}{g_sl_s(2\pi l_s)^2}\Delta.
\end{eqnarray}

A few words about the reparametrization are in order.
One might think that the rotation is just the gauge artifact that could be removed by a different gauge choice. However, this is not the case. Even when we use the following static gauge, 
\begin{eqnarray}\label{static}
T=\bar{t},\quad X=\bar{x},\quad \Phi=\bar{\varphi},
\end{eqnarray}
to make $\Phi_{\bar{t}}=0$, the mechanical motion cannot be removed away. The dynamical degree is just transfered to the radial profile. $R_{\bar{t}}\ne 0$ in general, because $R_t=R_{\bar{t}}+R_\varphi \Phi_t=0$. Owing to the reparametrization symmetry, the above Lagrangian (\ref{lagrangian}) remains the same but with the replacement of $-R_\varphi \Phi_t$ by $R_{\bar{t}}$. The BI fields of Eq. (\ref{BIfield}) will be also translated into the new language as $\bar{E}=E+\Phi_t B,\, \bar{B}=B$ and all the results below are valid in this new gauge too. However, we stress again that the gauge dependent statement about the magnetic flux (\ref{flux}) should be kept in mind. We still interpret the magnetic flux in the specific gauge (\ref{gauge}) as the net charge of the D0-branes, and the vanishing of the electric field $E$ ({\it not $\bar{E}$}) as the absence of the macroscopic IIA strings dissolved in the D2-brane. In fact, the two gauge dependent statements ($R_t=0$ and Eq. (\ref{BIfield})) define our system.

Since both the coordinate $\Phi$ and the BI gauge field component $A_x$ are dynamical, we have to consider their conjugated momenta   
\begin{eqnarray}\label{conjugate1}
\Pi=\frac{\partial{\cal L}}{\partial\Phi_t}=\frac{R^2\left[\Phi_t\left(R_\varphi^2+B^2\right)+EB\right]}{g_sl_s(2\pi l_s)^2\Delta}
\end{eqnarray}
and 
\begin{eqnarray}\label{conjugate2}
\Pi_A=\frac{\partial{\cal L}}{\partial E}=\frac{E\left(R^2+R_\varphi^2\right)+\Phi_t B R^2}{g_sl_s(2\pi l_s)^2\Delta}
\end{eqnarray}
to obtain the Hamiltonian,
\begin{eqnarray}\label{hamiltonian}
{\cal H}=\Pi\Phi_t+\Pi_AE-{\cal L}=\frac{R_\varphi^2+B^2+R^2\left(R_x^2+1\right)}{g_sl_s(2\pi l_s)^2\Delta}.
\end{eqnarray}

Varying the Lagrangian (\ref{lagrangian}) with respect to $\delta\Phi$ and $\delta A_{t,x,\varphi}$, one can obtain the equations of motion respectively as
\begin{eqnarray}
&&g_sl_s(2\pi l_s)^2\partial_t\Pi+\partial_x\frac{R^2R_xR_\varphi}{\Delta}+\partial_\varphi\frac{R^2\left(-1-R_x^2+E^2+\Phi_t B E\right)}{\Delta}=0,\label{eom1}\\
&&g_sl_s(2\pi l_s)^2\partial_x\Pi_A-\partial_\varphi\frac{E R_x R_\varphi}{\Delta}=0,\label{eom2}\\
&&g_sl_s(2\pi l_s)^2\partial_t\Pi_A+\partial_\varphi\frac{B\left(1-R^2\Phi_t^2\right)-E R^2\Phi_t}{\Delta}=0,\label{eom3}\\
&&\partial_t\frac{E R_xR_\varphi}{\Delta}+\partial_x\frac{B\left(1-R^2\Phi_t^2\right)}{\Delta}=0\label{eom4},
\end{eqnarray} 
where the second equation is the Gauss law constraint corresponding to the $U(1)$ local symmetry.

\section{Supersymmetry Analysis in General}\label{iii}
In this section, we perform the supersymmetry analysis explicitly. With an appropriate choice of $\Gamma$-matrix representation, we solve the Killing spinor equations in the general setup including the case where $E\ne 0$ and $\Phi_t\ne 0$. 
The number of supersymmetries preserved by the rotating tubular D2-brane is determined by the number of the Killing spinors $\epsilon$ satisfying
\begin{eqnarray}\label{killing}
\Delta\,\Gamma\,\epsilon&=&\left(\gamma_{tx\varphi}+E\gamma_\varphi\Gamma_{\natural}+B\gamma_t\,\Gamma_{\natural}\right)\epsilon\nonumber\\
&=&\left[R_\varphi\Gamma_{TXR}+R\,\Gamma_{TX\Phi}+RR_x\Gamma_{TR\Phi}+\Phi_t RR_\varphi\Gamma_{XR\Phi}\right.\nonumber\\
&&\left.+E\left(R_\varphi\Gamma_R+R\Gamma_\Phi\right)\Gamma_{\natural}+B\left(\Gamma_T+R\Phi_t\,\Gamma_{\Phi}\right)\Gamma_{\natural}\right]\epsilon\nonumber\\
&=&\Delta\,\epsilon,
\end{eqnarray}
where $\Gamma$ is the matrix defining kappa transformation on the world-volume of D-branes \cite{bergshoeff} and $\gamma_{\alpha}$'s and $\Gamma_{X^\mu}$'s are gamma matrix components in the world-volume coordinates and in the spacetime coordinates respectively. The operator $\Gamma_\natural$ is the chiral operator. Making use of the Killing spinor expression $\epsilon=\exp{(\Phi\Gamma_{R\Phi}/2)}\epsilon_0$ adapted for the embedding flat spacetime, one can obtain the following two independent equations;
\begin{eqnarray}
\left[R\left(R_x\Gamma_T+\Phi_t R_\varphi\Gamma_X\right)\Gamma_{R\Phi}+B\Gamma_T\Gamma_\natural-\Delta\right]\epsilon_0=0,&&\nonumber\\
\left[\left(R_\varphi\Gamma_R+R\,\Gamma_\Phi\right)\Gamma_{TX}+\left(E+\Phi_t B\right)R\,\Gamma_\Phi\Gamma_\natural+ER_\varphi\Gamma_R\Gamma_\natural\right]\epsilon_0=0,&&
\end{eqnarray} 
where $\epsilon_0$ is a constant 32-component Majorana spinor.
In order to work out these Killing spinor equations, it is convenient to recast them in the rectangular coordinates as
\begin{eqnarray}
\left[\Phi_t RR_\varphi\Gamma_{XYZ}+RR_x\Gamma_{TYZ}+B\Gamma_T\Gamma_\natural-\Delta\right]\epsilon_0=0,&&\nonumber\\
\left[\left(YR_\varphi-ZR\right)\Gamma_{TXY}+\left(ZR_\varphi+YR\right)\Gamma_{TXZ}\right.\qquad\qquad\qquad\qquad\qquad\qquad\qquad&&\nonumber\\
\left.+\left(EYR_\varphi-ZR\left(E+\Phi_t B\right)\right)\Gamma_Y\Gamma_\natural+\left(EZR_\varphi+YR\left(E+\Phi_t B\right)\right)\Gamma_Z\Gamma_\natural\right]\epsilon_0=0.&&
\end{eqnarray}

The {\bf s}-basis, which is well-illustrated in Ref. \cite{polchinski} and was applied in Ref. \cite{cho} for several intersecting D-branes, is a powerful tool to analyze the Killing spinor equations. 
The constant spinor $\epsilon_0$ can be written in this basis as
\begin{eqnarray}\label{abcd}
\epsilon_0&=&(a, b, c, d)\nonumber\\
&\equiv& a\,\, \vert +1,\, +1,\, 2s_2, 2s_3, 2s_4>+b\,\, \vert +1,\, -1,\, 2s_2, 2s_3, 2s_4>\nonumber\\
&+&c\,\, \vert -1,\, +1,\, 2s_2, 2s_3, 2s_4>+d\,\, \vert -1,\, -1,\, 2s_2, 2s_3, 2s_4>,
\end{eqnarray}
where the first and second entry $\pm1$'s denote the eigenvalues of $2S_0\equiv\Gamma^T\Gamma^X$ and $2S_1\equiv-i\Gamma^Y\Gamma^Z$ respectively and $2s_j$ are the eigenvalues of the operators $2S_j\equiv-i\Gamma^{2j}\Gamma^{2j+1},\,\,(j=2,3,4)$. Since the operators $2S_0$ and $2S_1$ do not commute with the operators of the above eigenspinor equations, the spinor $\epsilon_0$ must be some combination of all possible eigenstates of $2S_0$ and $2S_1$. With $\sigma\equiv\prod\limits_{j=2}^4(2s_j)$, the Killing spinor equations become
\begin{eqnarray}
i\Phi_t R R_\varphi(c, -d, a, -b)+iR R_x(-c, d, a, -b)-\sigma B(-c, d, -a, b)-\Delta(a, b, c, d)=0,&&\label{kse1}\\
\left(YR_\varphi-ZR\right)(-b, -a, d, c)+\sigma\left(EYR_\varphi-ZR\left(E+\Phi_t B\right)\right)(-b, a, d, -c)&&\nonumber\\
+i\left(ZR_\varphi+YR\right)(b, -a, -d, c)-i\sigma\left(EZR_\varphi+YR\left(E+\Phi_t B\right)\right)(-b, -a, d, c)=0.&&\label{kse2}
\end{eqnarray}
The former equation breaks supersymmetries by half, and the latter equation breaks another half.
Eq. (\ref{kse1}) has nontrivial solutions only when
\begin{eqnarray}
\Delta^2-\left(i\Phi_t R R_\varphi+\sigma B\right)^2-R^2R_x^2=0.
\end{eqnarray}
Therefore we obtain the necessary conditions for the supersymmetries;
\begin{eqnarray}\label{condition}
&&\Phi_t R_\varphi=0,\nonumber\\
&&R^2\left[1-\left(E+\Phi_t B\right)^2\right]+\left(1-E^2\right)R_\varphi^2=0,
\end{eqnarray}
for which Eq. (\ref{kse1}) relates coefficients as 
\begin{eqnarray}\label{relation}
\sigma B a=\sqrt{B^2+R^2R_x^2}\,c,\quad 
\sigma B b=\sqrt{B^2+R^2R_x^2}\,d,
\end{eqnarray}
which implies that half supersymmetries are broken by Eq. (\ref{kse1}) so far as the coefficients are constant. This is possible when $B=B_0RR_x$ with some constant $B_0$, if $R_x\ne0$, and otherwise, the magnetic field $B$ can take an arbitrary value. For these supersymmetric cases, $\Delta$ is simplified as
\begin{eqnarray}
\Delta=\sqrt{B^2+R^2 R_x^2}=\left\{
\begin{array}{ll}
R\vert R_x\vert\sqrt{B_0^2+1}&\quad(R_x\ne0)\cr
\vert B\vert&\quad(R_x=0)
\end{array}
\right.
\end{eqnarray}
It is easy to see from Eq. (\ref{kse2}) that for the states satisfying (\ref{condition}), the coefficients $a$ and $c$ vanish if $\sigma=E+\Phi_t B$, while $b$ and $d$ vanish if $\sigma=-(E+\Phi_t B)$. Hence $4+4=8$ symmetries are preserved. Now let us check the equations of motion, (\ref{eom1})-(\ref{eom4}) upon the imposition of the supersymmetric conditions.  

\subsection{Non-rotating Case}\label{iiia}

More specifically, when $\Phi_t=0$, the second condition of Eq. (\ref{condition}) requires $E^2=1$. Without loss of generality, we let $E=1$, then Eq. (\ref{kse2}) have eight solutions, four of which are of the form $(a, 0, \mbox{sgn}(B)a, 0)$ with $\sigma=1$ and the other four spinors take the form $(0, b, 0, -\mbox{sgn}(B)b)$ with $\sigma=-1$. 

i) Especially if $R_x=0$, the equations of motion (\ref{eom2})-(\ref{eom4}) dictate that $\Pi_A$ may depend only on $\varphi$, and $\mbox{sgn}(B)$ is constant over the tube world-volume. Therefore 8 supersymmetries are preserved for the tubular D2-branes with arbitrary cross sections (that is $R_\varphi\ne0$). The relation between the radial profile and the conserved charges $\Pi_A$ and $B$ is provided by Eq. (\ref{conjugate2}); $R^2+R_\varphi^2=g_sl_s(2\pi l_s)^2\vert\Pi_A\vert\vert B\vert$.   Hence the results of the circular supertubes \cite{townsend} and the supertubes with arbitrary cross sections of Ref. \cite{mateos} are recovered and the radial profile agrees well with the specific result about the elliptic supertube case discussed in Ref. \cite{cho1}. 

ii) If $R_x\ne0$, then $B=B_0R R_x$ so that the coefficients $a, b, c,$ and $d$ in Eq. (\ref{relation}) be constant. From the equation of motion (\ref{eom3}) we see that, in order for the conjugate momentum $\Pi_A$ to be conserved, the $B$ field ought to keep its signature along the angular direction of the `tube'. In view of Eq. (\ref{eom4}), it is reasonable to assume that the field $B$ does not change its signature over the D2-brane world-volume, in other words, there is no abrupt orientation flip on the world-volume. (This relation between the signature of the $B$ field and the orientation of the D2-brane was shown in detail in Ref. \cite{cho} for several supersymmetric configurations of D-branes.) The radial profile $R(x, \varphi)$ can then be obtained by solving the Gauss law constraint (\ref{eom1}): 
\begin{eqnarray}\label{govern}
\partial_x\left(\frac{1+\left(\partial_\varphi\ln{R}\right)^2}{\partial_x\ln{R}}\right)=\partial_\varphi^2\ln{R}.
\end{eqnarray}
When $R_\varphi=0$, it was shown in Ref. \cite{townsend} that the solution describes the two dimensional version of BIon spike of Ref. \cite{callan} carrying the magnetic flux. In general case, it is not easy to solve this nonlinear equation. However, we emphasize that there might be more general class of solutions for the `supertube' with the radial profile $R(x, \varphi)$ satisfying the above equation.

\subsection{Rotating Case}\label{iiib}

When $\Phi_t\ne0$, supersymmetry conditions (\ref{condition}) require $R_\varphi=0$ and $(E+\Phi_t B)^2=1$. The former condition is very interesting because it implies that the rotating supertube can take only the circular shaped cross section. This result is in contrast with the situation of the non-rotating case discussed previously. As we will see below, this is related with the fact that the tube is no longer tensionless if its cross section takes an arbitrary shape. In the latter condition, the combination $E+\Phi_t B$ replaces the role of the electric field in the non-rotating supertube case. The rotation of the magnetic sources (D0-branes) induces the electric field $\Phi_t B$. In view of the equations of motion, (\ref{eom1}) and (\ref{eom3}), we reasonably assume that $\partial_\varphi B=\partial_\varphi\Phi_t=0$ to achieve the conservation of the momenta, $\Pi$ and $\Pi_A$. Upon the imposition of the supersymmetric conditions \footnote{Without loss of generality we may set $E+\Phi_t B=1$.}, the Gauss law constraint (\ref{eom2}) and Eq. (\ref{eom4}) reduce to  
\begin{eqnarray}
\partial_x\frac{R^2}{\sqrt{B^2+R^2 R_x^2}}=0, \qquad
\partial_x\frac{B\left(1-R^2\Phi_t^2\right)}{\sqrt{B^2+R^2 R_x^2}}=0.
\end{eqnarray}

i) For the tubular case ($R_x=0$), these equations constrain the magnetic field $B$ and the angular velocity $\Phi_t$ to be $x$-independent. The radius of the supertube is determined by the conserved charges as
\begin{eqnarray}\label{radius}
R^2=g_sl_s(2\pi l_s)^2\vert \Pi_A B\vert=g_sl_s(2\pi l_s)^2\vert\Pi\vert.
\end{eqnarray}

ii) As for the non-tubular case ($R_x\ne0$), the above equations determine the radius $R(x)$ and the angular velocity $\Phi_t$ as
\begin{eqnarray}
R(x)=C\exp{\left[x/e_0\right]},\qquad R\Phi_t=v_0,
\end{eqnarray}
where $C, e_0$ and $v_0$ are integration constants. One very exciting thing is that the BI field strength becomes now rephrased in the spacetime coordinates $(T, R, \Phi)$ as
\begin{eqnarray}
F&=&E\,dt\wedge dx+B\,dx\wedge d\varphi\nonumber\\
&=&E\,dT\wedge\frac{e_0}{R}dR+\frac{Be_0}{R}dR\wedge\left(d\Phi-\Phi_t dT\right)\nonumber\\
&=&\left(E+\Phi_t B\right)\frac{e_0}{R}dT\wedge dR+B_0dR\wedge Rd\Phi.
\end{eqnarray}
Since the supersymmetric condition always requires $E+\Phi_t B=1$ regardless of the value of $E$, the solution describes a BIon spike on a D2-brane with a constant magnetic field $B_0$, {\it even when $E=0$.} The spike plays the role of the electric source with the charge $e_0$. The whole configuration carries inhomogeneous angular momentum about the axial direction;
\begin{eqnarray}
\vert\Pi\vert=\frac{\vert B_0\vert R^2}{g_sl_s(2\pi l_s)^2\sqrt{1+B_0^2}}.
\end{eqnarray}
This angular momentum is due to the mechanical rotation of the whole configuration with constant linear speed $R\,\Phi_t=v_0$ at every point. (See Fig. 1. for the case when $E=0$.) Although the BIon spike produces the electric field of the charge $e_0$, one can no longer interpret this as a fundamental string ending on a D0-charged D2-brane. The above solution is valid even when $E=0$, in which case the field momentum $\Pi_A$ is totally transferred to the mechanical momentum $\Pi$ via their relation (\ref{radius}). We will discuss about its interpretation later.

\epsfbox{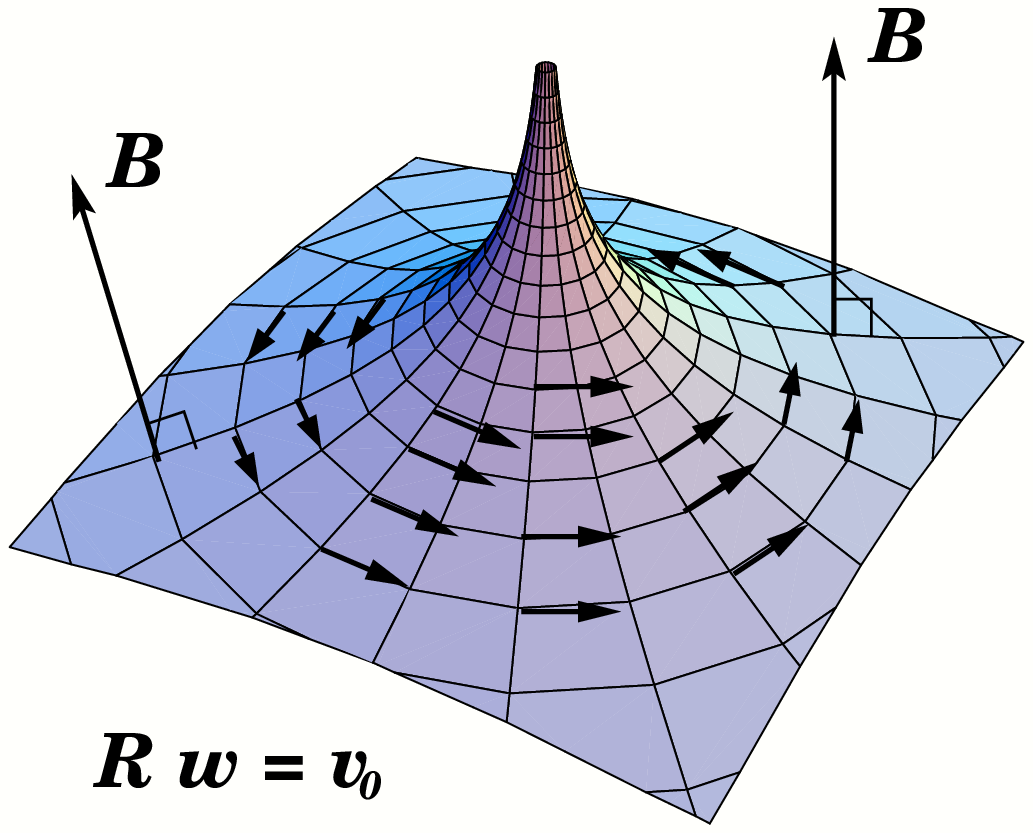}
\begin{flushleft}
{\footnotesize Fig. 2: The figure shows the BIon spike solution especially when BI electric field $E=0$. The whole configuration carries the angular momentum due to the mechanical rotation, which triggers the electric charge $e_0$ for the BIon. The linear speed is constant $v_0$ everywhere over the configuration.
}
\end{flushleft}

\section{Tensionless Brane?}\label{iv}

In the rotating supertube case, the supersymmetric condition, $R_\varphi=0$, forbids the solution to have an arbitrary cross section, which is in contrast with the non-rotating supertube case. A simple way to confirm this is to check the tension along the angular direction (as was done in Ref. \cite{mateos}). The spacetime stress-energy tensor can be obtained by the well-known Schwinger method;
\begin{eqnarray}
T^{\mu\nu}(\bar{X})&=&\left.\frac{2}{\sqrt{-\det{G}}}\frac{\delta S}{\delta G_{\mu\nu}(\bar{X})}\right|_{G=\eta}\qquad\qquad(\mu,\nu=0,\cdots 9)\nonumber\\
&=&-\frac{1}{g_sl_s(2\pi l_s)^2}\int\limits d^3\sigma\sqrt{-\det\left(g+F\right)}\left(g+F\right)^{-1(\alpha\beta)}\partial_\alpha X^\mu\partial_\beta X^\nu\,\,\delta^{(10)}\left(X(\sigma)-\bar{X}\right)\nonumber\\
&\equiv&\int\limits d^3\sigma{\cal T}^{\mu\nu}\left(X(\sigma)\right)\,\,\delta^{(10)}\left(X(\sigma)-\bar{X}\right).
\end{eqnarray}

Making use of the relation (\ref{gauge}) between the spacetime coordinates and the world-volume coordinates, we obtain the stress-energy density as
\begin{eqnarray}\label{SEtensor}
{\cal T}^{TT}&=&\frac{1}{g_sl_s(2\pi l_s)^2\Delta}\left[R^2\left(1+R_x^2\right)+R_\varphi^2+B^2\right]\nonumber\\
{\cal T}^{TX}&=&\frac{R^2R_xR_\varphi\Phi_t}{g_sl_s(2\pi l_s)^2\Delta}\nonumber\\
{\cal T}^{T\Phi}&=&\frac{1}{g_sl_s(2\pi l_s)^2\Delta}\left[B\left(E+\Phi_t B\right)+\Phi_t R_\varphi^2\right]\nonumber\\
{\cal T}^{XX}&=&-\frac{1}{g_sl_s(2\pi l_s)^2\Delta}\left[R^2+R_\varphi^2\left(1-R^2\Phi_t^2\right)\right]\nonumber\\
{\cal T}^{X\Phi}&=&\frac{R_xR_\varphi}{g_sl_s(2\pi l_s)^2\Delta}\nonumber\\
{\cal T}^{\Phi\Phi}&=&\frac{1}{g_sl_s(2\pi l_s)^2\Delta}\left[\left(E+\Phi_t B\right)^2-1-R_x^2+\Phi_t^2R_\varphi^2\right].
\end{eqnarray}
Note in the above that all the $\Phi_t$-dependent terms come with $R_\varphi$ upon the imposition of the supersymmetric condition $E+\Phi_t B=1$. Therefore these terms disappear when we impose the supersymmetric condition $R_\varphi=0$. Especially ${\cal T}^{\Phi\Phi}$ component that describes the tension along the angular direction, vanishes only for the tubular D2-brane with circular cross section and $(E+\Phi_t B)^2=1$. Consequently, tensionless condition agrees with the supersymmetric conditions (\ref{condition}). If the tube violates either of these conditions, it is no longer tensionless and is not supersymmetric. 

\section{Fundamental Strings induced from Moving Flux}\label{v}

From the stress-energy tensor, one can see what the rotating supertube is composed of. Under the supersymmetric conditions (\ref{condition}), the components ${\cal T}^{TT}$, ${\cal T}^{T\Phi}$, and ${\cal T}^{XX}$ survive. The component ${\cal T}^{T\Phi}$ concerns the energy flow around the tube, which is entirely due to the rotation of D0-branes dissolved on the tube. This is because, the rotation of a pure tubular D2-brane (without any D0-brane on its world-volume) is not physical and can be regarded as a world-volume reparametrization. The component ${\cal T}^{XX}=-\Pi_A$ ($=-\Pi/B$, hereafter, in case when $E=0$) is difficult to understand. Unlike the non-rotating case, the component survives even when $E=0$ and $\Pi_A=0$ correspondingly. It is not from the contribution of D0 gas on the world-volume, since they behave like tensionless dust. It is also very difficult to conceive a tubular D2-brane that has tension only along one direction. Therefore we inevitably have to consider some one dimensional object as its source. The only candidate for this, in this type-IIA theory, will be the fundamental string. And then where does it come from (especially when $\Pi_A=0$)? Here we recall that the fundamental string is polarized when it moves in the magnetic field background \cite{susskind}. The fundamental strings living on the tubular D2-brane will be polarized into long strings as the BI flux rotates. The whole effect is the induced electric field by the movement of the magnetic flux.
For the supersymmetric cases, the component ${\cal T}^{TT}$, being the Hamiltonian ${\cal H}$ in Eq. (\ref{hamiltonian}), should be sum of some charges. Let us see the Hamiltonian for the rotating case ($\Phi_t\ne 0$ and $R_\varphi=0$):  
\begin{eqnarray}
{\cal H}=\left\{
\begin{array}{ll}
\vert\Pi_A\vert+\frac{1}{g_sl_s(2\pi l_s)^2}\vert B\vert&\qquad (R_x=0)\cr
\vert\Pi_A\vert+\frac{1}{g_sl_s(2\pi l_s)^2}\frac{\sqrt{1+B_0^2}}{\vert B_0\vert}\vert B\vert&\qquad (R_x\ne 0)
\end{array}
\right.
\end{eqnarray}
Therefore it takes the same form as in the non-rotating case, but now, $\vert\Pi_A\vert$ should be understood as the charge density of the fundamental strings composed of the `bare' strings or the `induced' strings.

\epsfbox{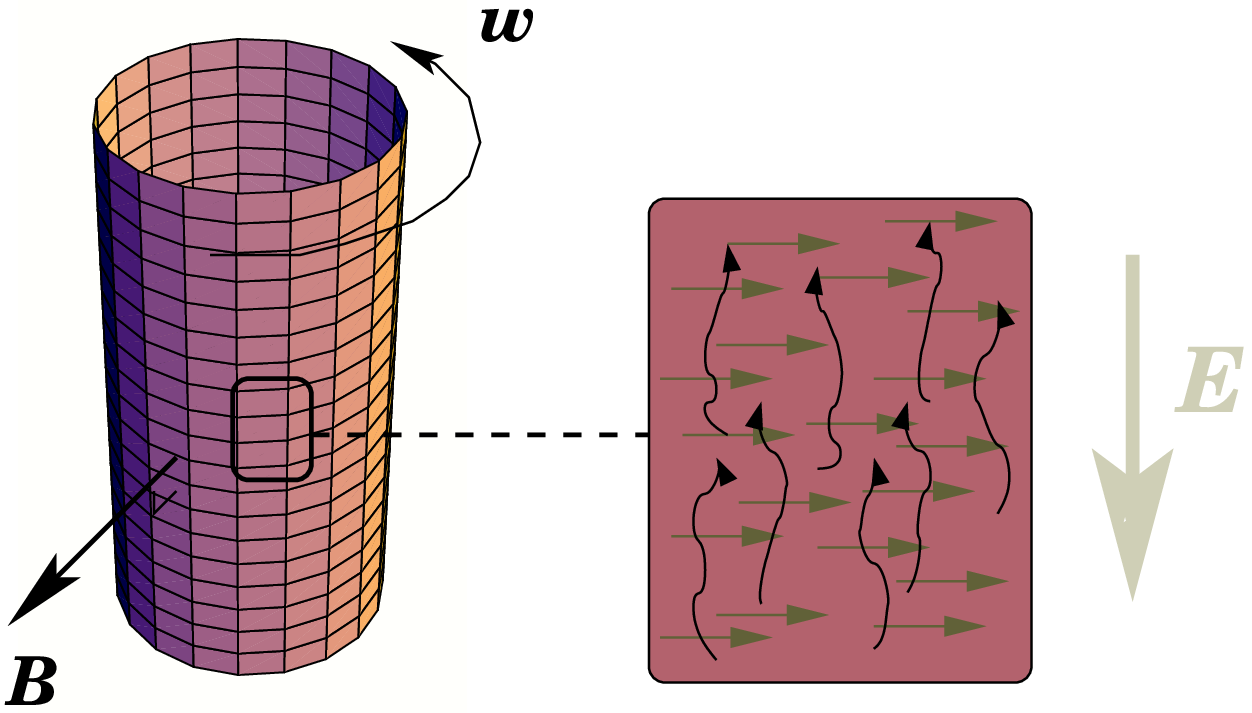}
\begin{flushleft}
{\footnotesize Fig. 3: As the D0-branes dissolved in the tube rotate (gray arrows), the fundamental strings (black wiggly arrows) living on the tube are polarized, which results in the electric field induction.
}
\end{flushleft}

\section{Angular Momentum Bound}\label{vi}

As for the rotating case, we may expect the angular momentum bound, otherwise the tangential linear speed of the dissolved D-particles will become superluminal. This is indeed the case. Making use of the supersymmetric condition $E+\Phi_t B=1$ and the radius $R$ in (\ref{radius}), one can easily obtain the following bound:
\begin{eqnarray}
\vert\Pi_A\vert=\left|\frac{\Pi}{B}\right|\leq\frac{1}{g_sl_s(2\pi l_s)^2}\frac{\vert B\vert}{\left(1-E\right)^2}.
\end{eqnarray}
When $E=0$, the bound shows that the charge density of the `induced' open strings cannot exceed that of D0-branes and the maximal tube radius is $\vert B\vert$.

In the non-rotating limit ($\Phi_t\rightarrow 0$), the supersymmetric condition requires $E\rightarrow 1$, therefore the bound disappears \footnote{It is not to be confused with the angular momentum bound for the system composed of a supertube and D0-charged superstrings along its axis, which was discussed in \cite{townsend}. What we are discussing in this paper is the bound of $\vert\Pi\vert=\vert\Pi_A B\vert$.}. However, this is a bit confusing. Although the angular momentum $\Pi_A$ is not mechanical, the open strings living on the supertube will be exerted by the Lorentz force and move along the angular direction. This is because the open string end points are oppositely charged for the BI fields. Therefore the same logic applied to the moving open strings might give some bound on $\Pi_A$. Although we will not discuss this point further in this paper, it should be clarified elsewhere. 
  
\section{Spiraling D-Helix}\label{vii}

When $R_x=0$, the axial direction will acquire isometry. Let the direction is compactified on a circle of radius $\lambda$. Performing T-duality on the rotating supertube along its axial direction, one obtains the spiraling D-helix. This can be seen by inspecting the open string boundary condition as was done in Ref. \cite{choh}. Since the spacetime components of BI fields are
\begin{eqnarray}
F=B\,dX\wedge\left(d\Phi-\Phi_t dT\right)+E\,dT\wedge dX,
\end{eqnarray}
the type -IIA open sting on the tube is subject to the following boundary conditions:
\begin{eqnarray}
&&\left.\partial_\sigma T+\partial_\tau X\left(E+\Phi_t B\right)\right|_{\sigma=0,\pi}=0\nonumber\\
&&\left.\partial_\sigma X+\partial_\tau T\left(E+\Phi_t B\right)-\partial_\tau\Phi B\right|_{\sigma=0,\pi}=0\nonumber\\
&&\left.\partial_\sigma \Phi\, R^2+\partial_\tau X\, B\right|_{\sigma=0,\pi}=0.
\end{eqnarray}
Under T-duality along $X$-direction, the boundary conditions become
\begin{eqnarray}
&&\left.\partial_\sigma \left[\tilde{T}+\tilde{X}\left(E+\Phi_t B\right)\right]\right|_{\sigma=0,\pi}=0\nonumber\\
&&\left.\partial_\tau \left[\tilde{X}+\tilde{T}\left(E+\Phi_t B\right)-\tilde{\Phi}\, B\right]\right|_{\sigma=0,\pi}=0\nonumber\\
&&\left.\partial_\sigma \left[\tilde{\Phi}\, R^2+ \tilde{X}\, B\right]\right|_{\sigma=0,\pi}=0,
\end{eqnarray}
where $(\tilde{T}, \tilde{X}, \tilde{\Phi})$ denote the dual coordinates in type-IIB theory. The second condition defines the hypersurface $\tilde{X}+\tilde{T}\left(E+\Phi_t B\right)-\tilde{\Phi}\, B=$constant, which describes nothing but the D-string profile on which the dual string lives. In the same comoving gauge as was used in type-IIA case, that is, $t=\tilde{T}$ and $\varphi=\tilde{\Phi}-\Phi_t\tilde{T}$, one can easily see that the D-string is coiled to form a helix with the tilting angle
\begin{eqnarray}
\tan{\theta}=\frac{\partial\tilde{X}}{R\partial\varphi}=\frac{B}{R},
\end{eqnarray}
and moving up with the speed
\begin{eqnarray}
\frac{\partial\tilde{X}}{\partial t}=-E\equiv v_{\vert\vert}.
\end{eqnarray}

The Lagrangian is exactly the same as that of type-IIA case:
\begin{eqnarray}
{\cal L}_{IIB}=-\frac{1}{g'_{s}l_s(2\pi l_s)}\sqrt{R^2\left[1-\left(E+\Phi_t B\right)^2\right]+B^2},
\end{eqnarray}
where $g'_{s}=g_sl_s/\lambda$ is the string coupling constant of type-IIB theory.
Therefore the Hamiltonian is minimized at the same radius, $R^2=\vert\Pi\vert$. One would obtain the same supersymmetric condition as in type-IIA case;\begin{eqnarray}
E+\Phi_t B=E+\frac{B}{R}\cdot \left(R\Phi_t\right)=E+\tan{\theta}\cdot v_\bot=1.
\end{eqnarray} 
The latter part, $\tan{\theta}\cdot v_\bot$ describes the speed at which the pitch looks like moving down as the helix spirals, henceforth called as the `virtual speed'. The supersymmetric condition regulates that the sum of the actual speed and the virtual speed should be that of light. 

\epsfbox{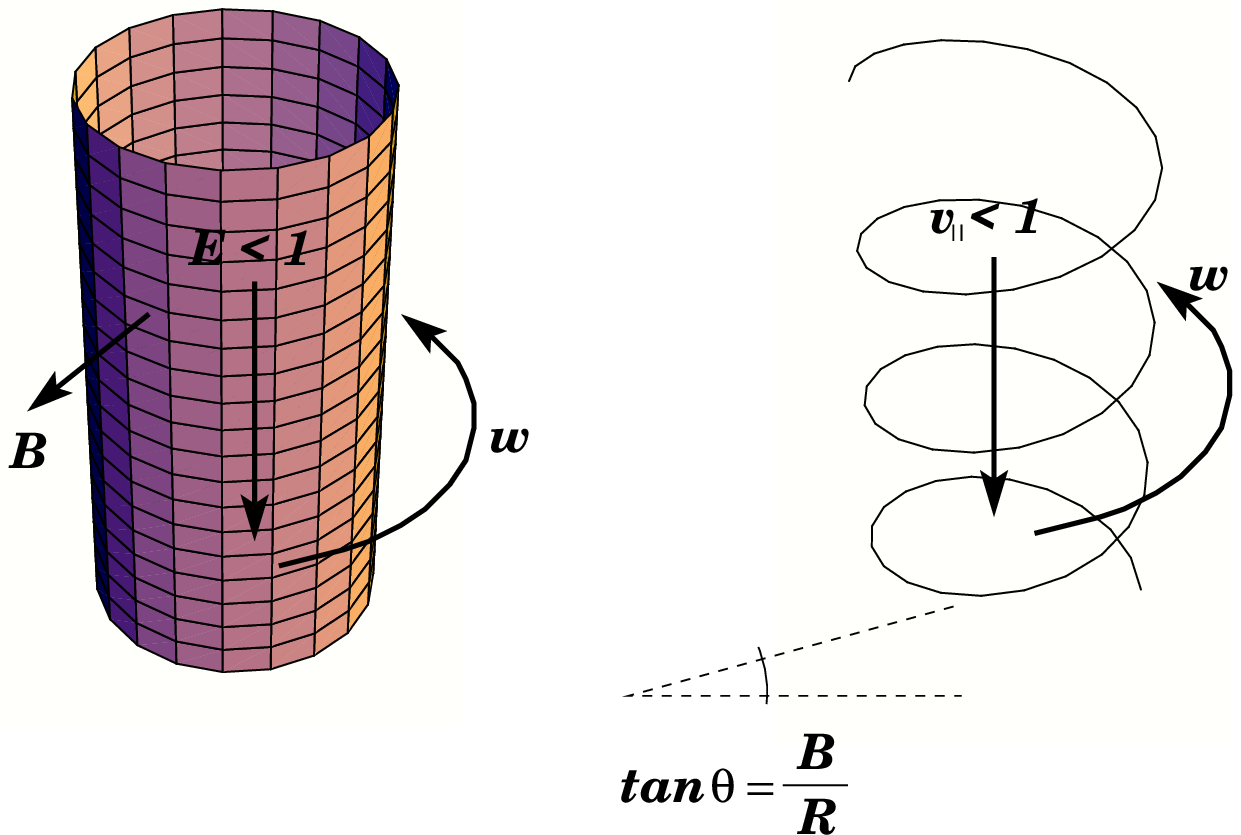}
\begin{flushleft}
{\footnotesize Fig. 4: The rotating supertube (Left) is T-dual to the spiraling super D-helix (Right). The charge density of the bare fundamental strings (therefore, winding) on the tube gives the momentum to the helix in the dual picture. The rotation of the tube corresponds to the spiral motion of the helix. The pitch of the helix is in a real motion due to the momentum and is, at the same time, in a virtual motion due the spiraling of the helix. The net speed (real plus virtual) amounts to the speed of light. 
}
\end{flushleft}

\section{Discussions}\label{viii}

In this paper, we considered the generalization of the supertube configuration of Ref. \cite{townsend} to the nonstatic case. We allowed the angular motion of an arbitrary shaped tubular D2-brane and checked whether it preserves any supersymmetry. 

When there is no angular motion ($\Phi_t=0$), a large class of radial profiles $R(x, \varphi)$ preserve 8 supersymmetries. These include the straight ($R_x=0$) tube with an arbitrary cross section, which was analyzed in \cite{mateos,cho1}. As for non-straight ($R_x\ne 0$) supertubes, we found the governing equation (\ref{govern}) for their radial profiles $R(x, \varphi)$. 

We showed that any rotational motion of a noncircular tubular D2-brane breaks supersymmetry completely. Only the circular shaped supertube can rotate preserving 8 supersymmetries. Although the rotational motion of the circular supertube is tangential to its world-volume, thus can be absorbed into a reparametrization, this motion is physical in the presence of magnetic field as in Eq. (\ref{BIfield}). Any motion of D0-branes is transverse to their world-volumes and gives rise to a term {\it proportional to the angular velocity} in the momentum flow ${\cal T}^{T\Phi}$. (See Eq. (\ref{SEtensor}).) Especially under T-duality along the axial direction of a rotating straight supertube, the rotational motion becomes the spiraling motion, which is transverse to the world-sheet of the resulting D-helix. The upshot is as follows. The electric field on a D2-brane is caused by the end points of superstrings \cite{townsend} while the magnetic flux  corresponds to the net RR-charge of D0-branes dissolved in the D2-brane \cite{townsend3}. Since the notion of the `electric' and `magnetic' fields is changed under the reparametrization transformation, we have to define them in some specific gauge. Hence this gauge choice determines the particular system. 

The whole physics depends on whether the BI field strengths (thus macroscopic strings and D0-branes) are static or rotating with respect to the spacetime. Had we worked in the static gauge of (\ref{static}) and defined the BI field strengths as $F=\bar{E}d\bar{t}\wedge d\bar{x}+\bar{B}d\bar{x}\wedge d\bar{\varphi}$, the problem would have changed into that of checking whether any supersymmetry-preserving radial motion ($R_{\bar{t}}\ne 0$) is allowed. The BI field strengths are given in the {\it static} world-volume coordinates, as in the ordinary supertube case \cite{townsend}. The answer can be inferred from our analysis and is that the supertube is possible only when $R_{\bar{t}}=0$ and $\bar{E}=1$. There is no constraint on the radial profile in this case.    

Let us look at the M-brane configuration corresponding to the rotating supertube case ($R_x=0$). The M2-brane action constructed from the D2-brane action shows the following relation between the M2-brane profile and the BI fields over the D2-brane \cite{schmidhuber,alwis,bergshoeff2}:
\begin{eqnarray}
\partial_\alpha X^{11}=\frac{1}{2}\frac{g_{\alpha\beta}\epsilon^{\alpha\beta\gamma}F_{\beta\gamma}}{\Delta}.
\end{eqnarray}

Since the metric (\ref{metric}) induced on the supertube is stationary rather than static, the motion of the corresponding M2-brane becomes
\begin{eqnarray}\label{motion}
\partial_tX^{11}=-\mbox{sgn}(B)+g_sl_s(2\pi l_s)^2\vert\Pi_A\vert \Phi_t.
\end{eqnarray}
Therefore the motion along $X^{11}$-axis is due to D0-branes on the one hand and due to the rotation of the tube on the other. The M2-brane extends to the $x$-coordinate, i.e., $\partial_xX^{11}=0$ and is coiled to form a `M-ribbon' \cite{ohta} as 
\begin{eqnarray}
\partial_\varphi X^{11}=g_sl_s(2\pi l_s)^2\vert \Pi_A\vert.
\end{eqnarray}
Interestingly, the rotating motion of the supertube $d\Phi=d\varphi+\Phi_t dt$ has been traded to the linear motion along $X^{11}$ (\ref{motion}) via the relation,
\begin{eqnarray}
X^{11}=-\mbox{sgn}(B)T+g_sl_s(2\pi l_s)^2\vert\Pi_A\vert \Phi.
\end{eqnarray}

The supersymmetric condition $R_\varphi=0$ can be understood in a different view point. If ever the rotating supertube admits arbitrary cross sections, one could deform the tube to make a D2/$\overline{\mbox D2}$ pair as was done in Ref. \cite{bak}. By taking T-duality along $x$-axis, one could get two D-strings tilted at the angle $\theta$ and $\pi-\theta$ respectively and passing by each other with the same speed. (See Fig. 2.) The resulting configuration looks very similar to the scissors configuration discussed in \cite{bachas,myers,cho}. However, it is not supersymmetric for any values of the tilting angle and the speed (even for the null scissors). See Appendix for details. This confirms that the rotating supertube does not admit arbitrary cross sections. 

A couple of remarks concerning future works are in order. The first one is on the zero radius limit of the rotating supertube. The zero radius limit of the non-rotating supertube corresponds to the D0-charged fundamental strings. The charge densities of the D0-branes and the fundamental strings are well encoded onto the tubular D2-brane as the magnetic flux and the electric displacement. As for the rotating supertube, the zero radius limit looks a bit ambiguous. Especially when $E=0$, the angular momentum of the tube is entirely due to the mechanical rotation of the tube. Although the `induced' strings play the role of the `bare' strings at any nonzero radius of the tube, it is unclear whether the `induced' strings make sense in the zero radius limit. It is rather likely that an array of D0-branes with net spin angular momentum is blown up to the rotating supertube. These points deserves further investigation for clarification. 

After the work on the supertube \cite{townsend}, many related $1/4$ supersymmetric configurations have been discovered \cite{peet,lee,park,lugo,maldacena,verlinde}. It would be interesting to generalize those configurations to include the mechanical angular momentum and to check their supersymmetric conditions. 

\bigskip

\acknowledgments
The authors thank S. Hyun, O.-K. Kwon, T. Mateos, J.-H. Park and H. Shin for helpful discussions and comments. This work is supported in part by KOSEF through Project No. R01-2000-000-00021-0.

\appendix

\section{Two Tilted D-strings Passing by Each Other}\label{a1}
The supersymmetry preserved by a tilted D-string that is in motion to the right is written as the combination of the left moving part and the right moving part of the supercharges on the open string world-sheet: 
$Q_{\alpha}+(\bar{\beta}^2\beta_2^\bot \tilde{Q})_\alpha$  \cite{polchinski}. Here 
\begin{eqnarray}
\beta_2^\bot=\prod\limits_{m=3}^9 \beta^m,
\end{eqnarray}
and $\beta^m=\Gamma^m\Gamma$ is the spacetime parity operator (the inversion along $x^m$-axis) on the world-sheet. The factor $\bar{\beta}^2=\rho(\gamma)\rho(\theta)\beta^2\rho(-\theta)\rho(-\gamma)$ tilts (by an angle $\theta$) and boosts (with the boost parameter $\gamma$) the D-string. Here $\rho(\gamma)=\exp{\left(i\gamma\Sigma^{01}\right)}$ and $\rho(\phi_i)=\exp{\left(i\phi_i\Sigma^{2i-1,2i}\right)}$ with the Lorentz rotation elements $\Sigma^{\mu\nu}=-\frac{i}{4}[\Gamma^\mu,\,\Gamma^\nu]$ of $SO(2k, 1)$ in the spinor representation. As for another D-string tilted at an angle $\pi-\theta$ and moving to the left with the same speed, the factor $\bar{\beta}^2$ should be replaced by the factor $\bar{\beta}'^{2}=\rho(-\gamma)\rho(\pi-\theta)\beta^2\rho(\theta-\pi)\rho(\gamma)$. The unbroken supersymmetries for both D-strings are determined by the 16 component spinors $\epsilon$ invariant under
\begin{eqnarray}\label{eigenop}
\left(\bar{\beta}^{2}\beta_2^{\bot}\right)^{-1}\bar{\beta}'^{2}\beta_2^{\bot}=2\cosh{\gamma}\sin{\theta}\left(\cosh{\gamma}\sin{\theta}+\Gamma^0\Gamma^1\sinh{\gamma}\sin{\theta}-\Gamma^1\Gamma^2\cos{\theta}\right)-1.
\end{eqnarray}
\hspace{1.5in}
\epsfbox{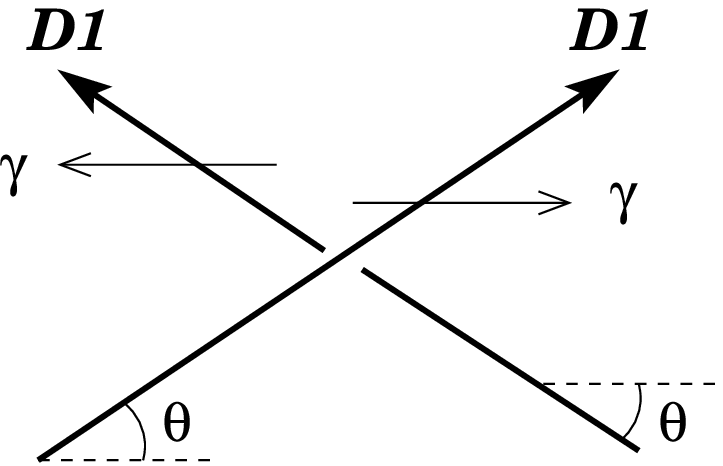}
\begin{flushleft}
{\footnotesize Fig. 5: Two tilted D-strings are passing by each other. This configuration is not supersymmetric in general. Only when the boost parameter $\gamma=0$ and the tilting angle $\theta=\pi/2$, the configuration preserve 16 supersymmetries. 
}
\end{flushleft}

As in Sec. \ref{iii}, the spinor $\epsilon$ is represented in the {\bf s}-basis as
\begin{eqnarray}\label{abcd1}
\epsilon&=&(a, b, c, d)\nonumber\\
&\equiv& a\,\, \vert +1,\, +1,\, 2s_2, 2s_3, 2s_4>+b\,\, \vert +1,\, -1,\, 2s_2, 2s_3, 2s_4>\nonumber\\
&+&c\,\, \vert -1,\, +1,\, 2s_2, 2s_3, 2s_4>+d\,\, \vert -1,\, -1,\, 2s_2, 2s_3, 2s_4>,
\end{eqnarray}
where the first and second entry $\pm1$'s denote the eigenvalues of $2S_0\equiv\Gamma^0\Gamma^9$ and $2S_j\equiv-i\Gamma^{2j-1}\Gamma^{2j}$ (a bit different from Sec. \ref{iii}) respectively and $2s_j$ are the corresponding eigenvalues which are subject to the Weyl condition $\prod_{a=0}^{4}2s_a=1$. The spinors $\epsilon$ invariant under the operator (\ref{eigenop}) satisfy
\begin{eqnarray}
&&\left(\cosh^2{\gamma}\sin^2{\theta}-1\right)(a,\,b,\,c,\,d)+\cosh{\gamma}\sinh{\gamma}\sin^2{\theta}(d,\,c,\,-b,\,-a)\nonumber\\
&&\qquad\qquad\qquad\qquad\qquad\qquad+i\cosh{\gamma}\cos{\theta}\sin{\theta}(a,\,-b,\,c,\,-d)=0.
\end{eqnarray}

This equation has nontrivial solutions for $a,\,b,\,c,\,d$, only when
\begin{eqnarray}
\left(\cosh^2{\gamma}-1\right)^2+\cosh^2{\gamma}\cos^2{\theta}\sin^2{\theta}+\cosh^2{\gamma}\sinh^2{\gamma}\sin^4{\theta}=0.
\end{eqnarray}
The only possible case to satisfy this condition is that $\gamma=0$ and $\theta=\pi/2$, that is when the two D-strings are parallel and at rest. For this case, $a,\,b,\,c,\,d$ are arbitrary, therefore 16 supersymmetries are preserved.


\end{document}